\documentclass[%
 reprint,
 amsmath,amssymb,
 aps,
]{revtex4-1}

\usepackage{graphicx}% Include figure files
\usepackage{dcolumn}% Align table columns on decimal point
\usepackage{bm}% bold math

\begin{document}

\title{Van der Waals density functional study of the structural and electronic properties of La-doped phenanthrene}

\author{Xun-Wang Yan$^{1,3,4}$}
\author{Zhongbing Huang$^{1,2}$}
\email{huangzb@hubu.edu.cn}
\author{Hai-Qing Lin$^1$}
\affiliation{$^1$Beijing Computational Science Research Center, Beijing 100084, China}
\affiliation{$^2$Faculty of Physics and Electronic Technology, Hubei University, Wuhan 430062, China}
\affiliation{$^3$State Key Laboratory of Theoretical Physics, Institute of Theoretical Physics,
Chinese Academy of Science, Beijing  100190, China }
\affiliation{$^4$School of physics and electrical engineering, Anyang Normal University, Henan 455000, China}
\date{\today}

\begin{abstract}
By the first principles calculations based on the van der Waals density functional theory, we study the crystal structures and electronic properties of La-doped phenanthrene.
Two stable atomic geometries of La$_1$phenanthrene are obtained by relaxation of atomic positions from various initial structures.
The structure-\uppercase\expandafter{\romannumeral1} is a metal with two energy bands crossing the Fermi level, while the structure-\uppercase\expandafter{\romannumeral2} displays
a semiconducting state with an energy gap of 0.15 eV, which has an energy gain of 0.42 eV per unit cell compared to the structure-\uppercase\expandafter{\romannumeral1}.
The most striking feature of La$_1$phenanthrene is that La $5d$ electrons make a significant contribution to the total density of state (DOS) around the Fermi level, which is
distinct from potassium doped phenanthrene and picene. Our findings provide an important foundation for the understanding of superconductivity in La-doped phenanthrene.
\end{abstract}

\maketitle

\section{Introduction}
%1 describe three experiments + chenxianghui Ba La experimet)
 The recent discovery of superconductivity with T$_c$ of 18 K in potassium doped picene (C$_{22}$H$_{11}$)~\cite{Mitsuhashi2010} has attracted great attention on the
 aromatic superconductor, a new class of organic superconductor, which was formed by intercalating alkali metal atoms into the interstitial space between aromatic molecules
 in the molecular crystal. Subsequently, alkali metal doped phenanthrene (C$_{14}$H$_{10}$)~\cite{Wang2011} and dibenzopentacene (C$_{30}$H$_{18}$)~\cite{Xue2012}
 with superconducting transition temperature of T$_c$ = 5.6~K and 33~K were reported, respectively. In the three kinds of superconductors, the electronic charges
 are transferred from the doped metal atoms to aromatic molecules and delocalized throughout the molecular crystal. Apart from alkali metal, alkaline earth and rare
 earth metal are also adopted to synthesize metal-doped phenanthrene crystal to explore their superconductivity. In experiment, the high quality
 Ba$_{1.5}$phenanthrene~\cite{Wang2011a} and La$_1$phenanthrene~\cite{Wang2012} superconductors were obtained with superconducting shielding fractions
 $40\%$ and $46\%$, and for La$_1$phenanthrene, the superconducting transition temperature significantly increases with the pressure and sustains high T$_c$ in a
 wide range of pressure~\cite{Chen2013}.

%2 describe the electronic structure papers  , theory)
 The determination of crystal structure of metal doped picene, phenanthrene, and dibenzopentacene is an important prerequisite to understand the electronic behavior
 and superconducting mechanism in the aromatic superconductors.
 However, the detailed crystal structures of doped materials have not yet been reported in experiment, especially the metal atomic
 position, due to the degradation of sample in air and the limit of measurement techniques~\cite{Wang2012}. Several theoretical groups have devoted
 to the researches on the crystal and electronic properties~\cite{Kosugi2011,DeAndres2011a,Kubozono2011,Cudazzo2011,Giovannetti2011,Verges2012,Huang2012,Roth2010}. For exmaple,
 Kosugi~{\it et al.} systematically investigated the crystal structure of K-doped picene with different K concentrations, and found that multiple structures exist
 at the same K concentration; Andres~{\it et al.} fully optimized the geometries of K$_3$picene and acquired two distinct structures,
 one with a herringbone structure
 and the other with a laminar structure.
For tripotassium-intercalated phenanthrene, there also exist the great discrepancy between the optimized structure and the measured one~\cite{DeAndres2011}.
%For tripotassium-intercalated phenanthrene, there also exists the same difficulty in determining the crystal structure~\cite{DeAndres2011}.
%  In general, different initial positions of six K atoms in the interstitial space between two picene or phenanthrene molecules can
% result in different final structures~\cite{DeAndres2011a, Kosugi2011}.
 %Moreover, the optimized cell parameters show a large discrepancy relative to the experimental measurements.
Other theoretical researches on the electronic correlation effect and magnetism of K$_3$picene were also performed and the lattice parameters were fixed at the measured values~\cite{Kim2011} in their first principle calculations.
 %Therefore, the determination of crystal structure of metal doped aromatic compounds
% is still a challenge and needs further experimental and theoretical studies.
%Additionally,

We notice that van der Waals interactions have not been taken into account in previous
 theoretical studies.
 The organic compounds of picene, phenanthrene, and dibenzopentacene are the typical molecular crystals, which have low density and hardness because of the light
 element and relatively long intermolecular bonds. In these materials, van der Waals interactions existing among molecules play a key role in determining the
 intermolecular distance and the stacking pattern. Therefore, it is quite necessary to include this factor in simulation of the structural and electronic properties
 of metal-doped picene, phenanthrene, and dibenzopentacene superconductors.

In this paper, we focus on the structural and electronic properties of rare earth metal La doped phenanthrene,
which became a valuable research issue because of the following reasons.
%emphasize the La is different from K.
Firstly, potassium is an alkali metal element with one electron in the outermost shell. Due to the high metal reactivity, potassium is usually used to intercalate into graphite, C$_{60}$, phenanthrene, picene, and dibenzopentacene to induce  superconductivity. Lanthanum belongs to rare earth metal with an electron configuration [Xe]5d$^1$6s$^2$ and exhibits two oxidation states, +3 and +2 in general. The metal reactivity of Lanthanum is much lower than potassium in the reactivity series of metal. Because of such difference, a large discrepancy relative to K-doped phenanthrene can be expected in La-doped phenanthrene.
%4 refer to this paper and this paper goal)
Secondly, the recent report from X. J. Chen group indicated that La-doped phenanthrene displays a significant positive pressure effect of T$_c$ and a sustainable high critical temperature within a wide pressure range, which are unique superconducting features different from K-doped phenanthrene~\cite{Chen2013}.
% Second, the unit cell of La-doped phenanthrene contains two La atoms, which are far less than six K atoms in the unit
%cell of K$_3$phenanthrene. The number of possible arrangements for two La atoms in a unit cell is much less than the one in K$_3$phenanthrene.

%(3 point out var de waals interaction importance)

%Third, the structural and electronic properties of La$_1$phenanthrene remain unclear.
The paper is organized as follows. In
Sec.~\ref{method} the computational method is briefly described. In Sec.~\ref{result} the results obtained are described and discussed.
In Sec.~\ref{summary}, the conclusions are presented.
\begin{figure}.
\includegraphics[width=8.5cm]{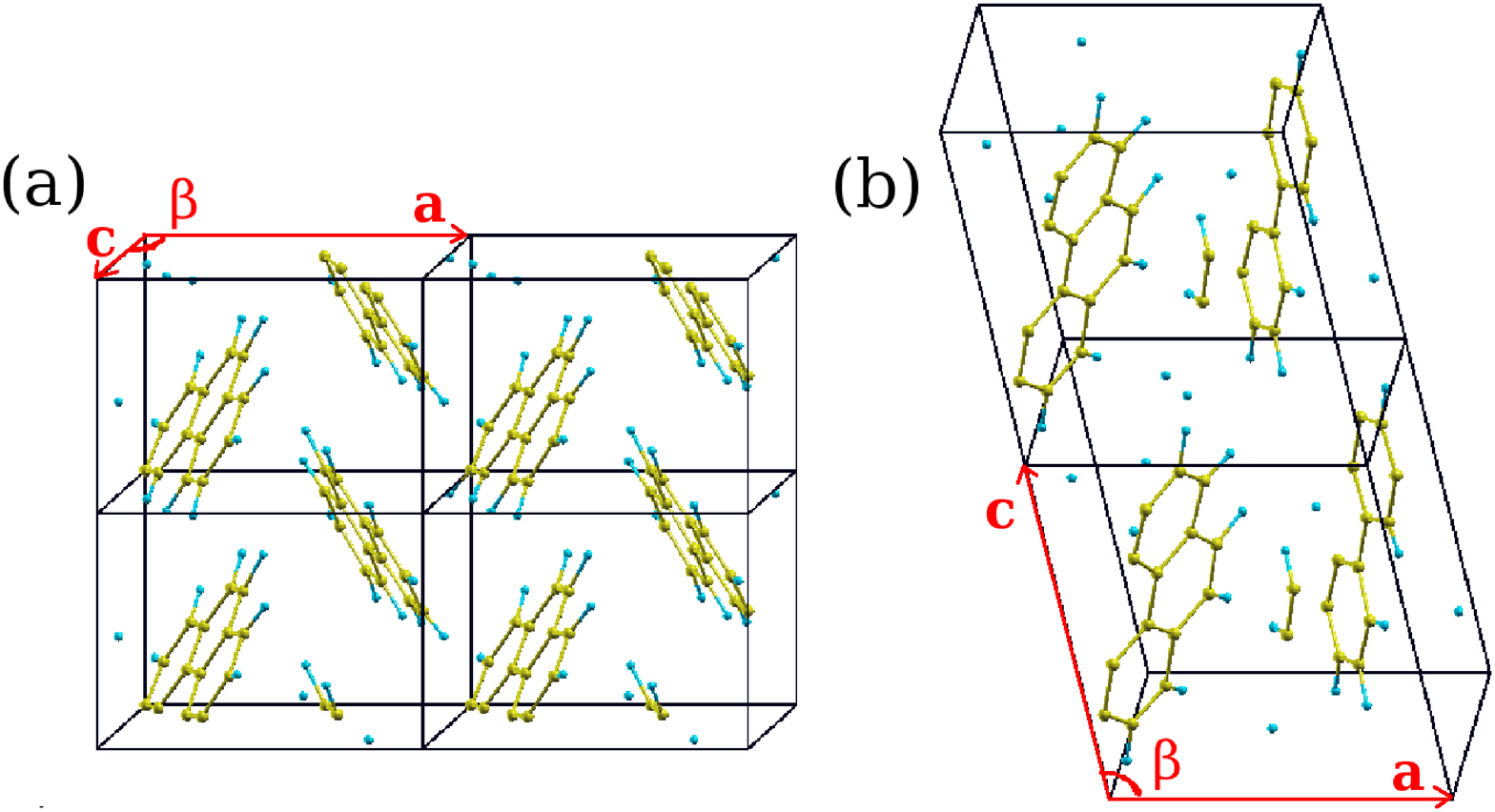}
\caption{(Color online)
The intralayer arrangement of molecules is shown in a 2 $\times$ 2 $\times$ 1 supercell (a) and the interlayer arrangement of molecules is shown
in a 1 $\times$ 1 $\times$ 2 supercell (b) for pristine phenanthrene. There are two phenanthrene molecules in a unit cell.} \label{geometry-no-La}
\end{figure}

\section{Details of Method}\label{method}
In our calculations the plane wave basis method was used~\cite{QE-2009}. We adopted the generalized gradient approximation
(GGA) with Perdew-Burke-Ernzerhof (PBE) formula~\cite{PhysRevLett.77.3865} for the exchange-correlation potentials. The projector augmented-wave method (PAW)
pseudopotentials were used to model the electron-ion interactions.
The local-density approximation (LDA) is appropriate for largely homogeneous systems, for example, simple metals and semiconductors. The semilocal-density approximation - GGA works well for inhomogeneous systems, for example, transition metals, ionic crystals, compound metals, surfaces, interfaces, and some chemical systems. However, for sparse systems, including soft matter, van der Waals complexes, and biomolecules, they have inter - particle separations, for which nonlocal, long-ranged interactions, such as van der Waals (vdW) forces, are influential.
The van der Waals interaction is considered in the scheme developed by Dion,
Thonhauser {\it et al.}~\cite{Dion2004,Thonhauser2007} and enclosed in the first principle calculation code {\it Quantum-ESPRESSO}~\cite{QE-2009}.
In the functional the exchange-correlation energy takes the form of
\begin{equation*}
E_{xc} = E_{x}^{GGA} + E_{c}^{LDA} + E_{c}^{nl}
\end{equation*}
where the exchange energy $E_{x}^{GGA}$ uses the revPBE GGA functional and $E_{c}^{LDA}$ is LDA to the correlation energy. $E_{c}^{nl}$ is the nonlocal energy term which accounts approximately for the nonlocal electron correlation effects. The vdW-DF method greatly improves the interaction energies of dispersion bonded systems \cite{Gulans2009a, Kelkkanen2009, Klimes2010}, and its application to hard solid can also obtain a systematic improvement in atomization energy and cohesive properties \cite{Klimes2011}.

 The Gaussian
broadening technique was employed and a mesh of $6\times 6\times 6$ k-points were sampled for the Brillouin-zone integration. After the full convergency test,
the kinetic energy cut-off and the charge density cut-off of the plane wave basis were chosen to be 80 Ry and 600 Ry, respectively.  In our calculations,
the lattice parameters were firstly fixed at the experimental values and the internal atomic coordinates were optimized by the energy minimization. The variable
unit cell calculations were performed to compare with the results in the case of fixed lattice constants.
%The convergence thresholds of the total energy,
%force on atom and pressure on cell are 10$^{-5}$ Ry, 10$^{-4}$ Ry/Bohr and 0.5 KBar respectively, which criteria are all satisfied in calculations.
The structure relaxation is performed by Broyden-Fletcher-Goldfarb-Shanno (BFGS) quasi-newton algorithm \cite{avriel2003, gilbert2006} and the maximum and minimum ionic displacement are set to 0.8 and 0.001 Bohr. The convergence thresholds of the total energy and force on atom are 10$^{-5}$ Ry and 10$^{-4}$ Ry/Bohr respectively, and for variable cell calculation, the residual stress on cell is less than 0.5 KBar.
\section{Results and Discussions}\label{result}
Pristine phenanthrene crystallizes in the space group $P2_{1}$ with the lattice parameters $a = 8.453 $ \AA, $b = 6.175$ \AA, $c = 9.477 $\AA, and
$\beta = 98.28 ^{\circ}$ measured at room temperature \cite{Wang2011}. The crystal geometry is shown in Fig. \ref{geometry-no-La}, where a unit cell contains two molecules arranged
in a herringbone structure. After the lattice parameters and the inner atomic coordinates are all fully relaxed, the obtained parameters $ a$, $b$, $c$,
and $\beta $ are 8.401 \AA, 6.166 \AA, 9.445 \AA, and 97.65$^{\circ}$, which are perfectly consistent with the ones in experiment.
If van der Waals interaction is not included in the calculation, the optimized parameters are 8.791 \AA, 6.368 \AA, 9.608 \AA, and 98.66$^{\circ}$.
The differences between two sets of computational cell parameters demonstrate that van der Waals force is an important aspect of physics in phenanthrene
and related compounds, and it is necessary to take account of this interaction in the first principles simulation for these materials.

\begin{figure}
\includegraphics[width=8.5cm]{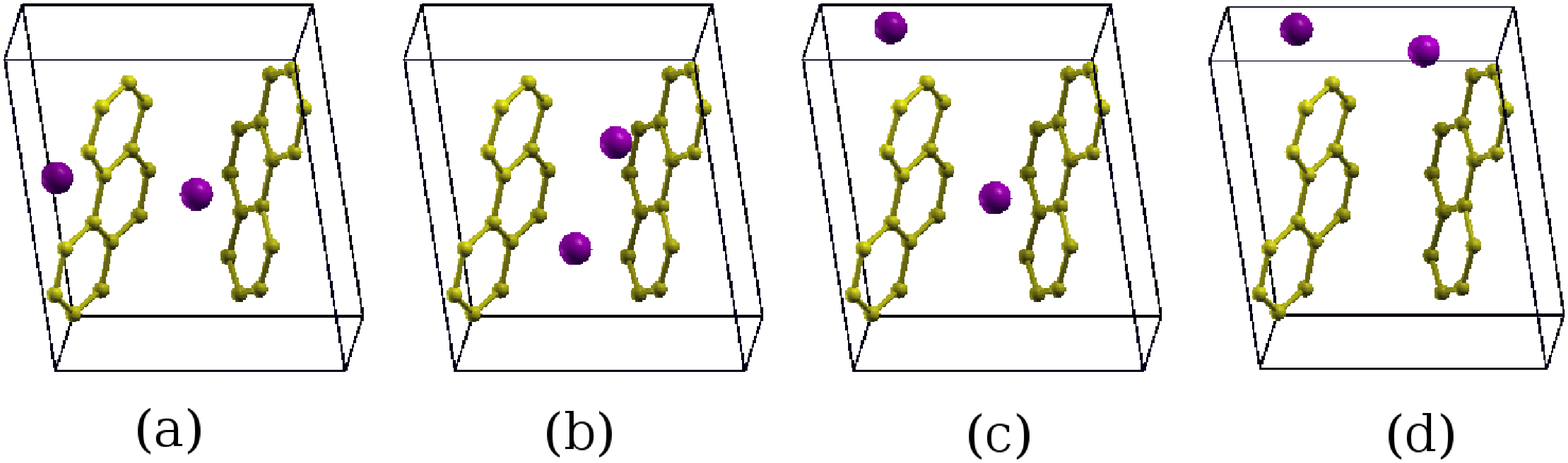}
\caption{(Color online)
Schematic diagram for the positions of two La atoms relative to two phenanthrene molecules. The four configurations are designed to start the structure relaxation.
Hydrogen atoms are not shown.} \label{La-initial-position}
\end{figure}

The rare earth metal La atoms are intercalated into phenanthrene solid with the ratio of La atom and molecule 1:1 to form La$_1$phenanthrene.
%The space group of doped phenanthrene crystal is also $P2_{1}$ and the lattice parameters have a small change relative to pristine phenanthrene~\cite{Wang2011}.
La-doped phenanthrene crystallizes in P$2_{1}$ space group with the lattice parameters a =8.482 \AA, b =6.187 \AA, c =9.512\AA, and $\beta$ =97.95$^{\circ}$ \cite{Wang2012}.
In the relaxation calculations, we have checked most possible configurations to explore the reasonable geometry structure.
For La$_1$phenanthrene, the possible arrangements for two La atoms and two molecules in a unit cell are much reduced compared to the situation of
K$_3$phenanthrene with six K atoms and two molecules in a unit cell. The following four cases were considered and believed to be sufficient for the
arrangement of molecules and La atoms, as shown in Fig.~\ref{La-initial-position}: (a) two La atoms are located at two sides of a molecule in a unit cell;
(b) two La atoms are located at the same side of a molecule; (c) one La atom sits at the end of molecule (between two molecular layers) and the other at
the side of molecule (in the molecular layer); (d) two La atoms sit at the molecular ends. We relaxed the inner atomic positions in the unit cell with cell
parameters fixed at the measured ones. The initial configurations (a), (c) and (d) would evolve into the same structure, called the structure-\uppercase\expandafter{\romannumeral1},
whereas the configuration (b) leads to the structure-\uppercase\expandafter{\romannumeral2}.

\subsubsection{La-doped phenanthrene: the structure-\uppercase\expandafter{\romannumeral1}}
\begin{figure}
\includegraphics[width=8.5cm]{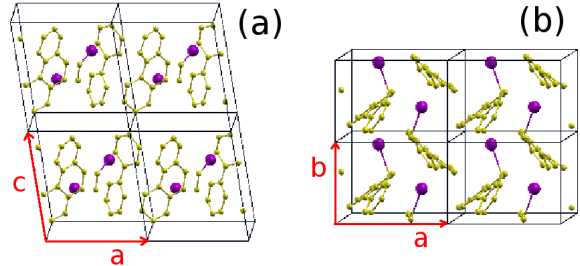}
\caption{(Color online)
Crystal structure of La$_1$phenanthrene for the structure-\uppercase\expandafter{\romannumeral1} with a La atom inserted in each hole. La atomic positions are viewed from two different angles in panels (a) and (b). Hydrogen atoms are not shown.
 } \label{geometry-1La}
\end{figure}

%0, geometry structure
Fig.~\ref{geometry-1La} shows one of the two stable structures, namely the structure-\uppercase\expandafter{\romannumeral1}. In a molecular layer of La-doped phenanthrene,
the aromatic molecules are arranged in a herringbone structure, with an angle about 90$^\circ$ between nearest-neighbor molecular planes. The arrangement of molecules
forms a network structure, and La atoms enter into the net hole. For the structure-\uppercase\expandafter{\romannumeral1}, La atoms are distributed more uniformly, i.e.
one La atom sited in each hole. The positions of two La atoms in a unit cell are about $\frac{1}{3}c$ and $\frac{2}{3}c$ along the $c$ axis.
As for the symmetry of unit cell, the structure-I also keeps P$2_{1}$ group symmetry similar to pristine phenanthrene, namely all atomic positions obey the symmetry operation (x, y, z) $\to$ (-x, y+0.5, -z).
\begin{figure}
\includegraphics[width=8.0cm]{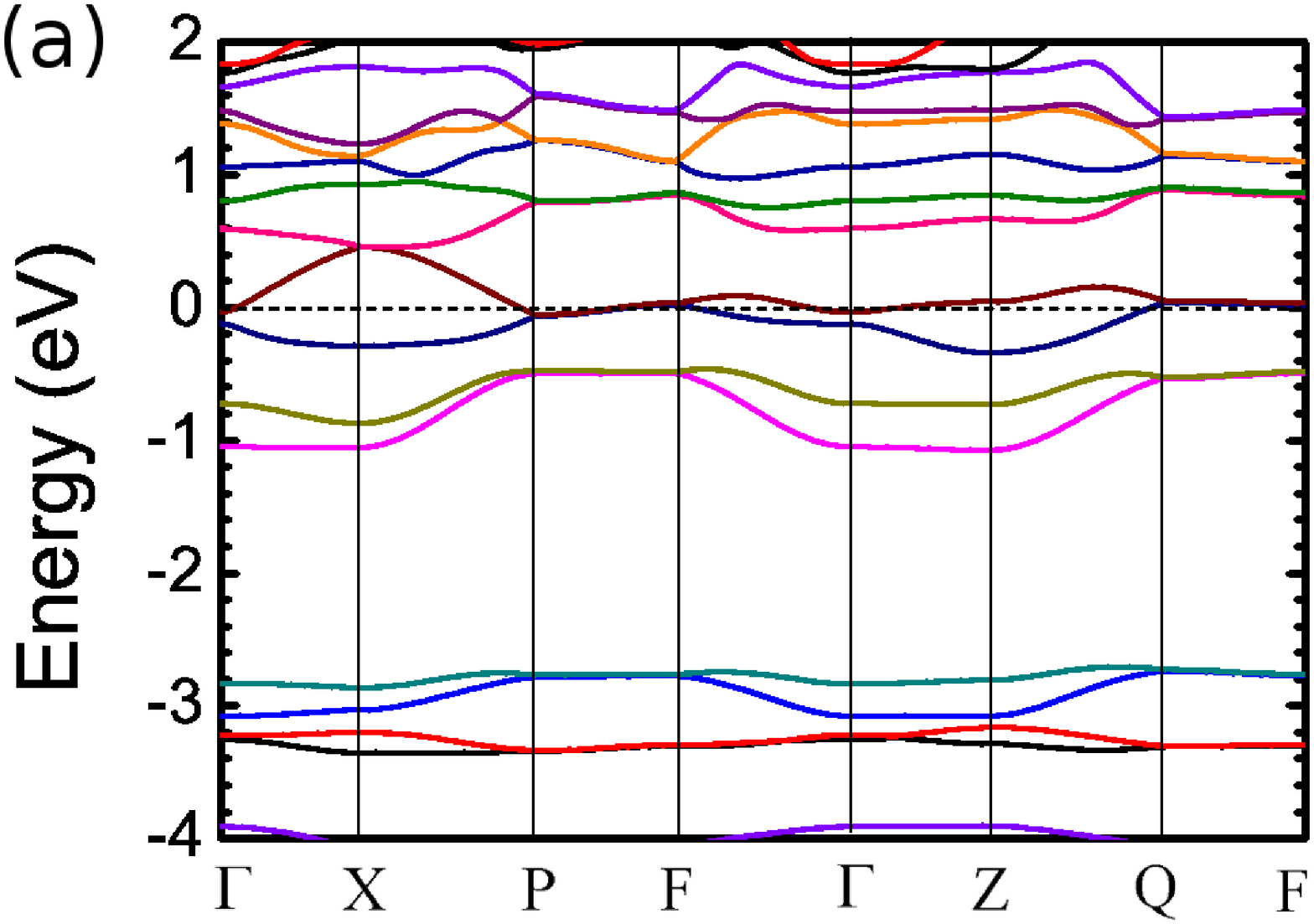}
\includegraphics[width=6.0cm]{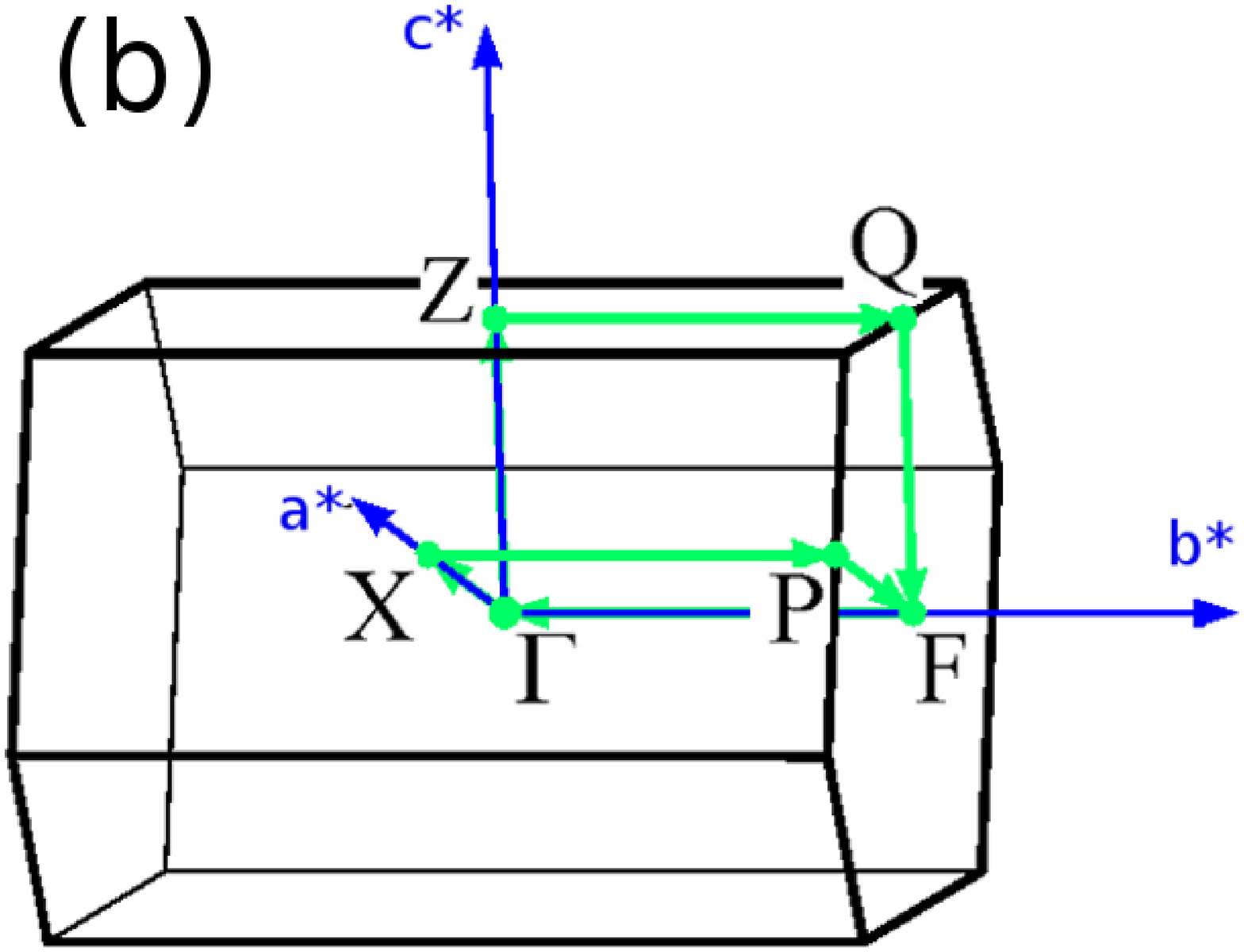}
\includegraphics[width=8.0cm]{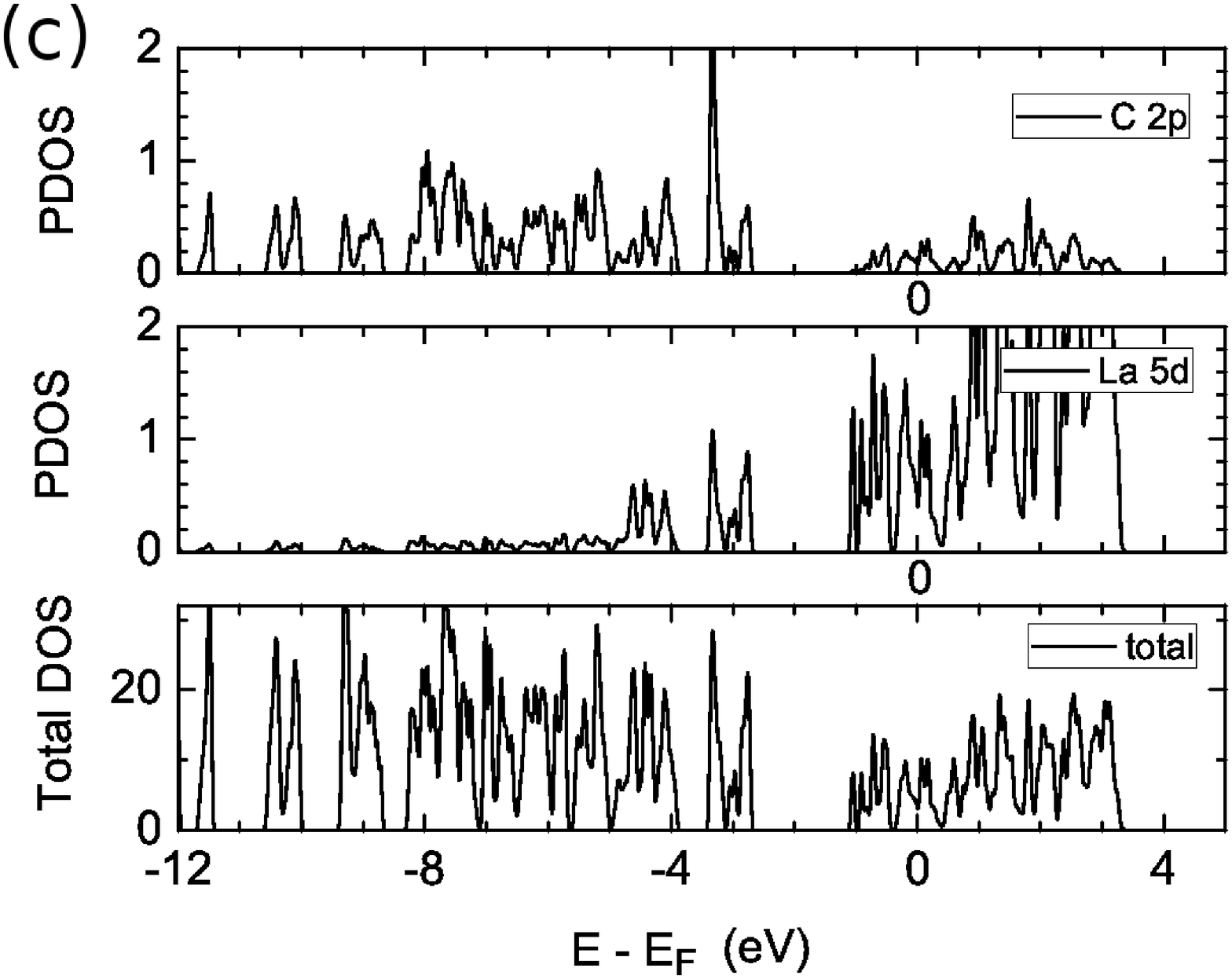}
\caption{(Color online)
(a) Band structure of La$_1$phenanthrene with experimental lattice parameters for the structure-\uppercase\expandafter{\romannumeral1}. (b) The Brillouin zone and selected $k$ point paths for energy band. (c) Total DOS per unit cell and orbital-resolved partial DOS per C atom and per La atom. The Fermi level is set to zero.
 } \label{1La-band-dos}
\end{figure}

%\begin{figure}
%\includegraphics[width=8.0cm]{1La-perhole-press-dos.eps}
%\caption{(Color online) Total DOS of La$_1$phenanthrene for the structure-\uppercase\expandafter{\romannumeral1} with different cell volumes are presented. $95\%$ V$_\mathrm{E}$, $90\%$ V$_\mathrm{E}$, and $85\%$ V$_\mathrm{E}$ indicate the contracted volumes, with V$_\mathrm{E}$ denoting the volume of unit cell at ambient pressure.
% } \label{1La-perhole-press-dos}
%\end{figure}

%1, first analysis the band and dos
Because of intercalation of La atom, the molecular plane of phenanthrene is distorted. Then the interactions among molecules, especially by the bridge of La atom, are enhanced and lead to a stronger band dispersion, as shown in Fig.~\ref{1La-band-dos}. Fig.~\ref{1La-band-dos}(a) clearly shows that there are two bands crossing the Fermi level.
The dispersion of bands along ${\varGamma}Z$ and $FQ$ (parallel to the axis $c^*$ in the Brillouin zone) is small, which is related to the weak interlayer coupling mediated by H-H overlap. In the direction of axis $b$, the La atoms and molecules are stacked most closely because of the short axis $b$, which is reflected in the stronger dispersion of bands along $XP$, $F\varGamma$ and $ZQ$ (parallel to the axis $b^*$).
Compared with the band structure of K$_3$phenanthrene in the lower panel of Fig.2 in Andres' paper \cite{DeAndres2011}, the total width of two bands crossing the Fermi level is about 0.8 eV
for La$_1$phenanthrene, which is larger than 0.5 eV in the K doped case.
In addition, the two bands below the Fermi level locate around -1.0 eV for La$_1$phenanthrene, lower than -0.5 eV in K$_3$phenanthrene.
%The reason for these differences is discussed in the following paragraphs.

An important feature of the electronic structure of La-doped phenanthrene is revealed in Fig.~\ref{1La-band-dos}(c) that besides the C $2p$ states, another important contribution to the DOS around the Fermi energy is from La $5d$ electrons with the weight of about 27$\%$, which is estimated based on the total DOS ($3.26 ~states/eV$) and projected DOS on La $5d$ orbitals ($0.44~ states/eV~ \times 2 ~for ~two ~La ~atoms$).

%2, analyze the L$\ddot{\text o}$wdin charge.
In order to understand this significant feature, we performed L$\ddot{\text o}$wdin population calculations to discover the charge transfer from La atom to phenanthrene molecule and resulting
electron occupations in atomic orbitals. The calculations indicate that each of two C atoms close to La atom gets about 0.35$\sim$0.40 electron from the neighbored H and La atoms. The $6s$ orbital
is empty and $5d$ orbital has 1.8 electron for La atom. As can be seen from Table. \ref{Lowdin}, the charge transferred from each La atom to a molecule is about 1.0 electron. This result has a great difference from the situation of K-doped phenanthrene. Three K atoms are required for each aromatic molecule in K-doped phenanthrene, picene, dibenzopentacene, coronene\cite{Kubozono2011}, as well as fullerene~\cite{RevModPhys.81.943,Iwasa2003}, where each molecule can get three electrons from metal atoms.
%So, La element was selected to intercalated into phenanthrene with the ratio of La and phenanthrene molecule 1:1, since La element can exhibit +3 valence state and is expected to provide three electrons to a molecule.
The K 4s states are far higher than the Fermi energy and completely empty in the K-doped system~\cite{DeAndres2011a}, while there exists strong hybridization between La 5d states and C 2p states in La$_1$phenanthrene, which can explain the difference between their band structures. The contribution of La $5d$ states to the DOS at the Fermi level implies that superconductivity in this compound
is intimately related to these La $5d$ electrons.
\begin{table}
\caption{\label{Lowdin} L$\ddot{\text o}$wdin population of La$_1$ phenanthrene in the structure-\uppercase\expandafter{\romannumeral1}. The number of valence electron in La pseudopotential is 11 and the total valence charge of phenanthrene molecule is 66. The spiling charge that can not be projected to the atomic orbital is 0.32.}
\begin{tabular}{c c c c }
\hline
\hline
&La ion&phenanthrene ion\\
Occupation &9.78 &  66.90 \\
\hline
& La 5s,6s & La 5p & La 5d  \\
Occupation &1.80 & 6.15 & 1.83  \\
\hline
\hline
\end{tabular}
\end{table}

%3 analysis the pressure
The positive pressure effect on T$_c$ is a common feature in the doped aromatic superconductors.
%Especially for La-doped phenanthrene, application of pressure not only significantly increases T$_c$ from its ambient-pressure value of 4.8 K to 12.3 K at 18.4 GPa but also stabilizes high T$_c$ over a wide pressure range.
Especially for La-doped phenanthrene, application of pressure significantly increases T$_c$ from 4.9 to 8.5 K as the pressure increases from ambient pressure to 3.2 GPa before the structural phase transition\cite{Chen2013}. A similar increase of Tc from 6.1 to 7.6 K with the pressure from ambient pressure to 1.0 GPa was also found by Wang et al.\cite{Wang2012}.
%To understand the effect of pressure on the superconductivity in La-doped phenanthrene, we keep the shape of crystal unit cell and decrease the volume to $95\%$, $90\%$, and $85\%$ to simulate the pressure-induced volume reduction. The corresponding pressures are 1.4 GPa, 3.3 GPa, and 6.0 GPa.
To understand the effect of pressure on the superconductivity in La-doped phenanthrene, we keep the shape of crystal unit cell and decrease the volume to simulate the pressure effect. When the volume changes from $100\%$ to $95\%$, $90\%$, and $85\%$, the corresponding pressure is increased from 2.27 to 3.61, 5.52 and 8.20 GPa, respectively.
The total DOS are displayed in Fig.~\ref{1La-perhole-press-dos} as a function of the volume.
%One can see that the DOS at the Fermi level increase with the volume decreasing.
The value of DOS at Fermi level increases gradually from 3.2 to 3.6, 3.9, 4.7 states/eV with the volume decreasing.
According to the Bardeen-Cooper-Schrieffer (BCS) theory of electron-phonon induced superconductivity, the increase of DOS with pressure will result in an increase of T$_c$, which is in consistent with the experimental measurements~\cite{Chen2013}.

\subsubsection{La-doped phenanthrene: the structure-\uppercase\expandafter{\romannumeral2}}
\begin{figure}
\includegraphics[width=8.5cm]{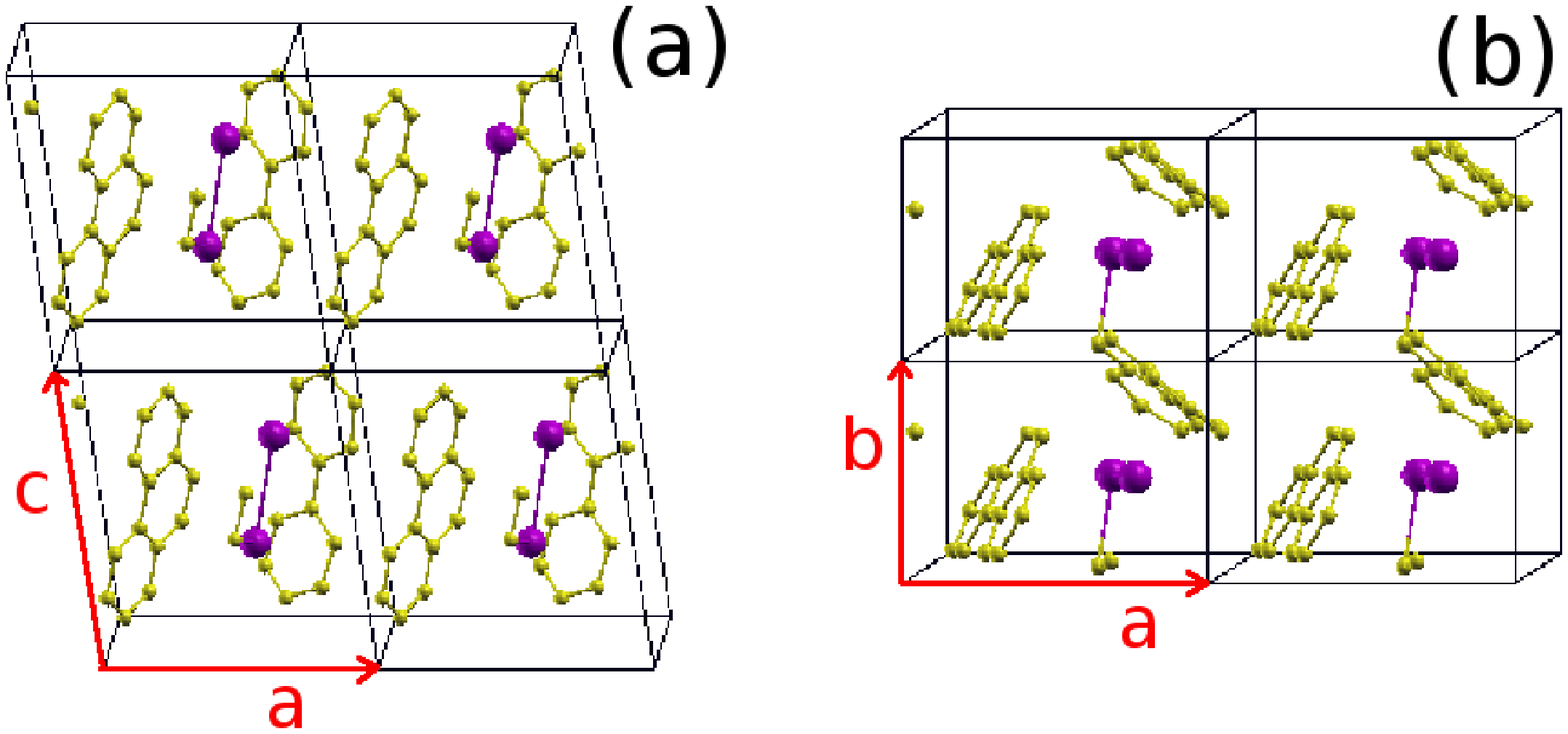}
\caption{(Color online)
Crystal structure of La$_1$phenanthrene for the structure-\uppercase\expandafter{\romannumeral2} with two La atoms inserted in the same hole. La atom positions are viewed from two different angles in panels (a) and (b). Hydrogen atoms are not shown.
 } \label{geometry-2La}
\end{figure}

For La$_1$phenanthrene, one unit cell includes two La atoms and two molecules. When two La atoms enter the same hole in the network formed by phenanthrene molecules, we obtain
the structure-\uppercase\expandafter{\romannumeral2}, as shown in Fig.~\ref{geometry-2La}.
Because two La atoms do not obey the symmetry operation (x, y, z) $\to$(-x, y+0.5, -z),
the structure-II does not keep the P$2_{1}$ group symmetry and it belongs to P1 space group.
%the optimized cell parameters are $a =$ 9.202 \AA, $b =$6.005 \AA, $c =$ 9.829 \AA, and $\beta =$103.66$^{\circ}$ compared with the experimental ones with the corresponding fractional errors of +8.5$\%$, -2.9$\%$, +3.3$\%$, and +5.8$\%$, respectively.
%From an energetic view of point, the structure-\uppercase\expandafter{\romannumeral2} is most stable among several configurations arranged in our calculations.
Compared with the structure-\uppercase\expandafter{\romannumeral1}, the energy is about 0.42 eV lower per unit cell composed of 50 atoms. Hence, the structure-\uppercase\expandafter{\romannumeral2}
is more stable from the energetic view.
Our results show that the stress of the structure-II with the experimental unit cell is 2.186 GPa, which is comparable to the stress of the structure -I and less than the pressure 3 GPa for tripotassium-intercalated phenanthrene \cite{DeAndres2011}.
\begin{figure}
\includegraphics[width=8.0cm]{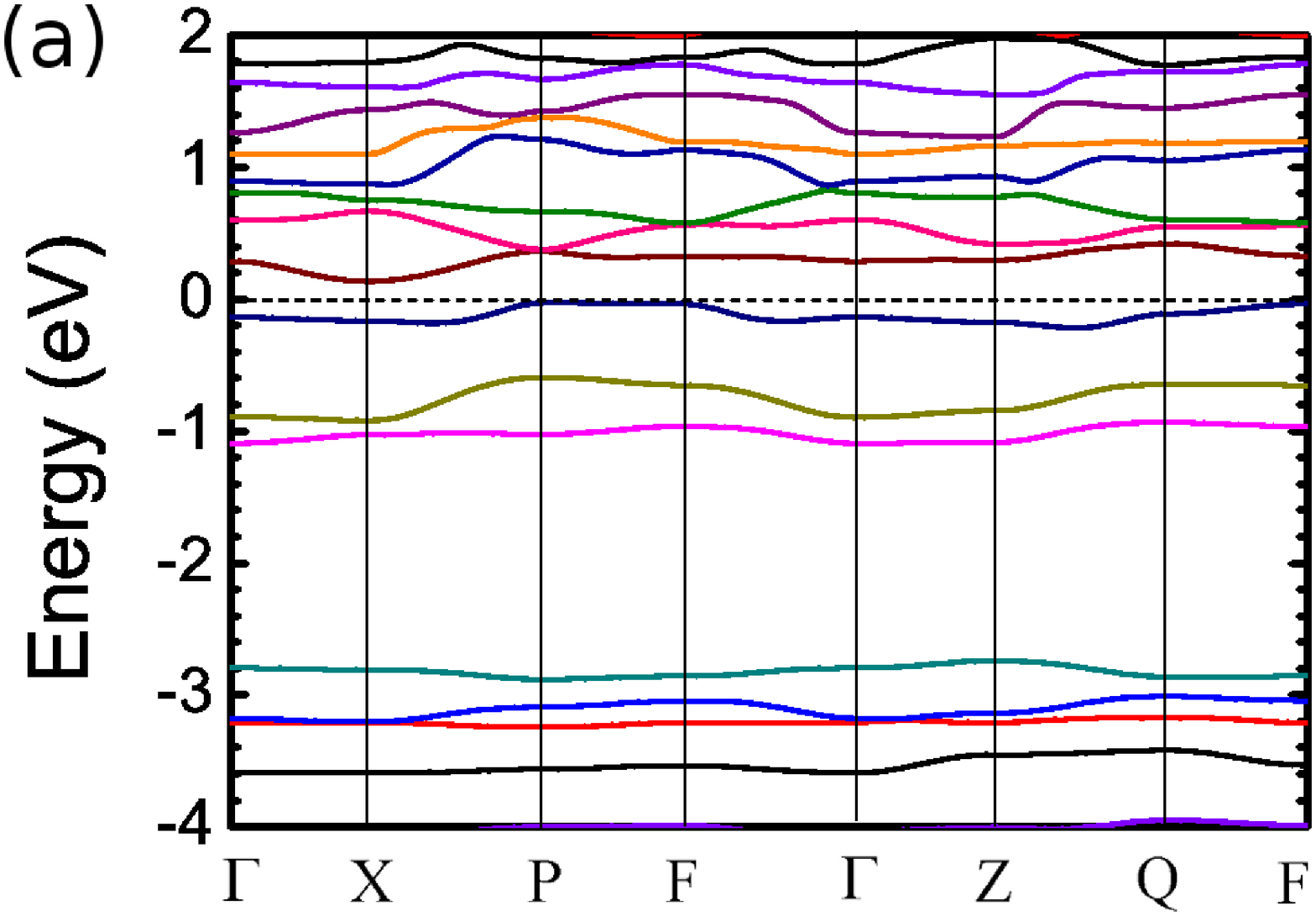}
\includegraphics[width=8.0cm]{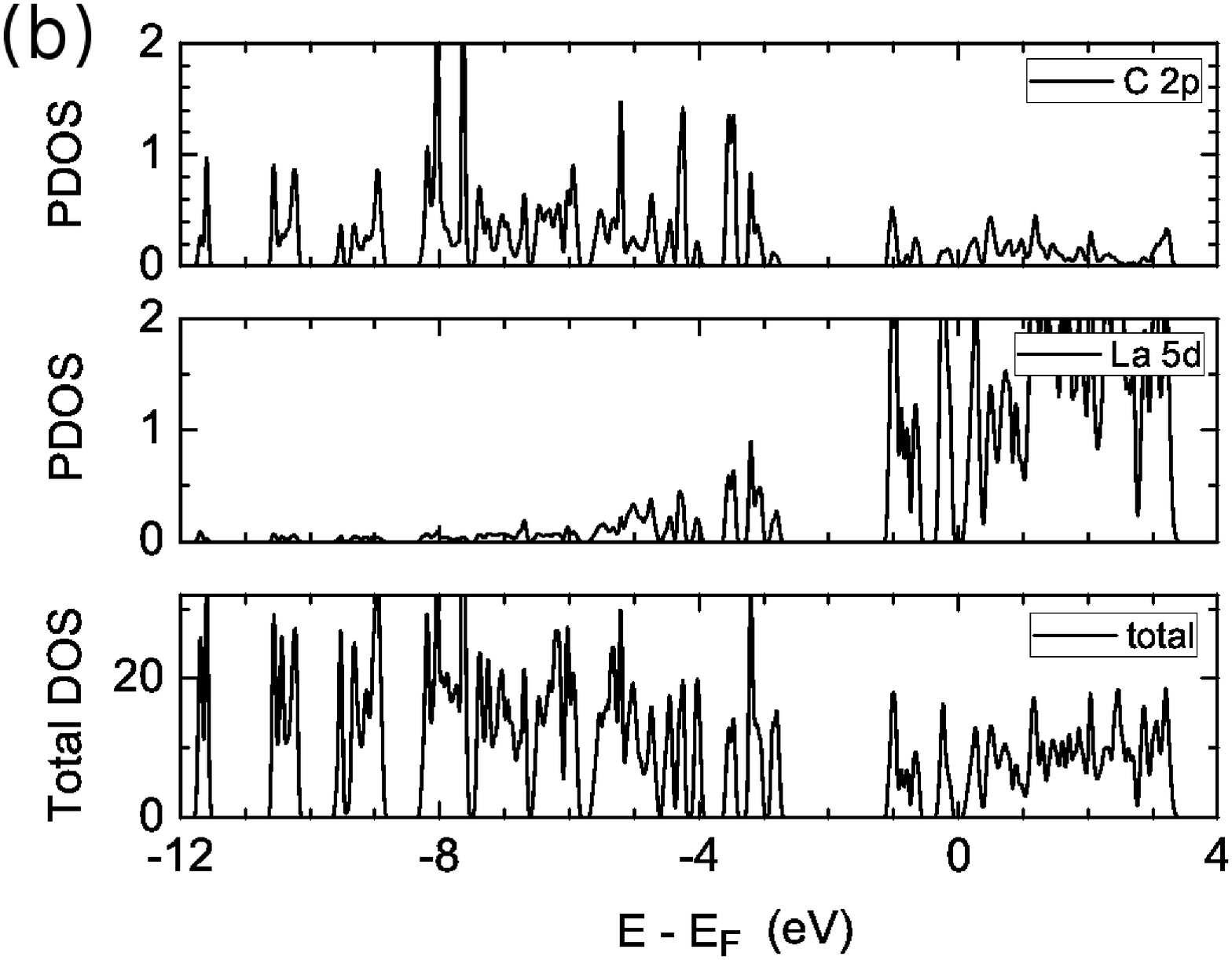}
\caption{(Color online)
(a) Band structure of La$_1$phenanthrene with experimental lattice parameters for the structure-\uppercase\expandafter{\romannumeral2}. (b) Total DOS per unit cell and orbital-resolved partial DOS per C atom and per La atom. The Fermi level is set to zero.
 } \label{2La-dos}
\end{figure}

Fig.~\ref{2La-dos} shows the band structure and DOS for the structure-\uppercase\expandafter{\romannumeral2}. The bands are more flat and less dispersive relative to the ones of the structure-\uppercase\expandafter{\romannumeral1}. An remarkable feature revealed by our calculations is that it displays a semiconducting state with an energy gap of 0.15 eV.
The lower panel in Fig.~\ref{2La-dos} shows the partial DOS of La $5d$ states, which make a great contribution to the total DOS around the Fermi level,
similar to the situation for the structure-\uppercase\expandafter{\romannumeral1}. L$\ddot{\text o}$wdin population calculations show that each of two C atoms close to La atom gets
about 0.35$\sim$0.40 electron from the neighbored H and La atoms, and La $6s$ orbital is empty and La $5d$ states have about 2.0 electron for La atom. The lost charge per La
atom is about 1.0 electron.
In experiment, the effect of charge transfer in La-doped phenanthrene was studied by Raman spectroscopy. Based on the similar behavior of phonon softening between La-doped and K-doped phenanthrene, X. F. Wang \textit{et al.} suggested that  three electrons are transferred from a La atom to a phenanthrene molecule\cite{Wang2012}.
Obviously, our L$\ddot{\text o}$wdin population analysis for the structure-\uppercase\expandafter{\romannumeral1} and structure \uppercase\expandafter{\romannumeral2} does not support their deduction. More experiments are needed for
clarifying the actual charge transfer in La-doped phenanthrene.

The semiconcuting structure-II phase is not the superconducting phase, but it may be a parent compound of superconductor when the concentration of La has few percents deviation from 1 in La$_1$phenanthrene, similar to the parent compounds of superconductor LaCuO$_4$ \cite{Muller1986}, KFe$_{1.5}$Se$_2$ \cite{Yan2011} and K$_{0.8}$Fe$_{1.6}$Se$_{2}$ \cite{Xun-WangYan}.

\subsubsection{The optimized lattice parameters for the structure-\uppercase\expandafter{\romannumeral1} and structure-\uppercase\expandafter{\romannumeral2}}
The simulation of crystal structure of doped phenanthrene or doped picene is a great challenge, because the optimized crystal structures of doped phenanthrene\cite{DeAndres2011} or doped picen\cite{Kosugi2011,DeAndres2011a} reported at present deviate largely from the measurements, even is completely different from the experiment.
The variable cell calculations were also performed for La-doped phenanthrene. For structure-\uppercase\expandafter{\romannumeral1} with P2$_1$ group symmetry, the optimized lattice parameters are $a =$8.632 \AA, $b =$ 6.505 \AA, $c =$ 9.661 \AA, and $\beta =$108.03$^{\circ}$. Compared with the experimental ones $a =$8.482 \AA, $b =$6.187 \AA, $c =$9.512 \AA, and $\beta =$97.95$^{\circ}$, the corresponding fractional errors are +1.7$\%$, +5.1$\%$, +1.5$\%$, and +10.3$\%$, respectively.
For the structure-\uppercase\expandafter{\romannumeral2} with P1 group symmetry, the calculated lattice parameters are $a =$ 9.202 \AA, $b =$6.005 \AA, $c =$ 9.829 \AA, $\alpha =$86.59$^{\circ}$, $\beta =$103.66$^{\circ}$ and $\gamma =$93.06$^{\circ}$, and the corresponding fractional errors are +8.5$\%$, -2.9$\%$, +3.3$\%$, -3.8$\%$, +5.8$\%$ and +3.4$\%$, respectively ($\alpha$ = $\gamma$ = 90$^{\circ}$ in the experimental cell with P2$_1$ symmetry).
The unit cell volumes for the structure-I and structure-II are 515.8 \AA$^3$ and 525.6 \AA$^3$, very close to the experimental value 494.3 \AA$^3$.
With the optimized lattice constants, we have investigated the electronic structure of La$_1$phenanthrene and found that their electronic properties are similar to ones with the experimental lattice parameters. The structure-\uppercase\expandafter{\romannumeral1} is a metal and the structure-\uppercase\expandafter{\romannumeral2} lies in the semiconducting state with a larger energy gap of 0.30 eV. Still, the DOS of La $5d$ orbital have a substantial contribution to the total DOS around the Fermi level.

There is very small difference of lattice parameters in experiment between pristine and La-doped phenanthrene. However, in our optimized
structure-\uppercase\expandafter{\romannumeral1} and structure-\uppercase\expandafter{\romannumeral2}, the errors for some parameters compared to the measurement are larger than 5$\%$.
One reason may be that in the compound containing rare earth metal La whose atomic radius is much larger than the ones of C and H atom, it is difficult to deal with the non-bonding
interaction between C atom and La atom with available pseudopotentials.

In experiment, the metallic phase of La$_1$phenathrene was observed, which corresponds to the structure-\uppercase\expandafter{\romannumeral1}.
Now the question arising is why the structure-\uppercase\expandafter{\romannumeral2} with lower energy was not observed in measurement?
This issue could be addressed in the relaxation calculations. As already mentioned before, the initial configurations (a), (c) and (d) in Fig.~\ref{La-initial-position}
evolve into the structure-\uppercase\expandafter{\romannumeral1}, while only the particular configuration (b) results in the structure-\uppercase\expandafter{\romannumeral2}.
This demonstrates that it is easier to form the metallic phase during the crystal growing process.

\section{Conclusions}\label{summary}
% summary
In summary, by the first principles calculations, we study the crystal structures and electronic properties of pristine and La-doped phenanthrene under considering van der Waals interaction.
For La$_1$phenanthrene, two stable atomic geometries are obtained with different La atom distributions in the interstitial space among molecules.
The structure-\uppercase\expandafter{\romannumeral1} is a metal and two energy bands cross the Fermi level.
The structure-\uppercase\expandafter{\romannumeral2} displays the semiconducting state with an energy gap of 0.15 eV.  From the energetic viewpoint, the structure-\uppercase\expandafter{\romannumeral2} is more stable because it has an energy gain of 0.42 eV per unit cell compared to the structure-\uppercase\expandafter{\romannumeral1}.
The most striking feature of La$_1$phenanthrene is that La $5d$ electrons make a great contribution to the total DOS around the Fermi level, which is obviously distinct
from potassium doped phenanthrene and picene. The similar results can be obtained when we carried out the related calculations with the optimized lattice parameters,
demonstrating that the electronic property of La$_1$phenanthrene is strongly correlated with the specific positions of La atoms.
Our findings provide an important information for understanding superconductivity in La$_1$phenanthrene.

Whether the superconductivity in these metal doped aromatic crystals is driven by electron-phonon or electron-electron mechanism is still in debate.
The value of $x=3$ is special for K$_x$picene, K$_x$phenanthrene and K$_x$dibenzopentacene,  indicating that three electrons doped into each molecule
seem to be a key factor to induce superconductivity in K-doped aromatic materials. For La$_1$phenanthrene, La atom is expected to provide three electrons
to each phenanthrene molecule. However, only about one electron is transferred from La atom to aromatic molecule, as indicated by L$\ddot{\text o}$wdin
population. This is drastically different from usual lanthanide compounds and lanthanide complexes, where La takes the valence state of +3 and each La
atom loses three electrons. Moreover, La $5d$ electrons have a great contribution to the DOS around the Fermi level, which might be intimately related to superconductivity,
even unconventional superconductivity.
%like $3d$ transition metal compounds LaOFeAs \cite{doi:10.1021/ja800073m} and LaCuO4\cite{LaCuO4}.
All these make La$_1$phenanthrene special and rich physical properties can be expected in further research.

\section{Acknowledgments}
We acknowledge the valuable discussion with Guo-Hua Zhong. This work was supported by MOST 2011CB922200, the Natural Science Foundation of China under Grants Nos.91230203, 91221103, 11174072 and U1204108.
%, and was also partially supported by the program of the State Key Laboratory of Theoretical Physics (No.Y3KF271CJ1), Institute of Theoretical Physics, Chinese Academy of Sciences.

{\it Note added.} While in the preparation of this manuscript, we learnt of a paper by S. Shahab Naghavi group (Phys. Rev. B 88, 115106 (2013)), which reports that the lowest energy state of La$_1$phenanthrene is band insulating and the best metallic structure is slightly higher in energy but retains the $P2_{1}$ symmetry. The structure-\uppercase\expandafter{\romannumeral1} and structure-\uppercase\expandafter{\romannumeral2} in our results are accordant with the best metallic structure and the lowest energy structure shown in Fig.~3 in Phys. Rev. B 88, 115106 (2013), respectively.

%\begin{references}
%\bibitem{Nature2010}
%Nature 464, 76-79 (4 March 2010)
%\bibitem{feng-naturem}
% Zhang Y, {\it et al.}, Nature Materials {\bf 10}, 273 (2011).
%\end{references}

%\bibliographystyle{apsrev4-1}
%\bibliography{Collection,revise-add,revise-add-2,Roth2010}
%\bibliography{}

\end{document}